\documentclass[aps,pra,twocolumn]{revtex4}
\usepackage{bbold}
\usepackage{graphicx}

\begin{document}

\title{Local control of remote entanglement}
\author{Peter \v{S}telmachovi\v{c}${}^{1,2}$  and Mari\'{a}n Ro\v{s}ko${}^{1,2}$}
\affiliation{
${}^{1}$ Research Center for Quantum Information, Slovak Academy of Sciences
, D\'{u}bravsk\'{a} cesta 9, 845 11 Bratislava, Slovakia \\
${}^{2}$ Quniverse, L{\'\i}\v{s}\v{c}ie \'{u}dolie 116, 841 04 Bratislava, Slovakia
}

\begin{abstract}
We address the problem of the generation of entanglement. We focus 
on the control of entanglement shared by  two non-interacting parties $A$ and $C$ via 
interaction with a third party $B$. We show that, for certain physical models, 
it is possible to have an asymptotically  complete control of the entanglement shared by $A$ and $C$ by
changing parameters of the Hamiltonian local at site $B$. 
We present an example where different models (propositions) of physical situation, 
 that lead to different descriptions of the system $B$, 
result into different amount entanglement produced. 
In the end we discuss limits of the procedure. 
\end{abstract}
\maketitle

\section{Introduction}

Quantum mechanics admits correlations  of a very specific type 
(entanglement) but the task to create  such correlations 
between several systems need not have a solution under given conditions.
A natural way how to correlate two systems is to use mutual (direct) interaction
between the two systems. Such approach is unusable where
the interaction is weak, the two systems are too far from each other, 
or simply they do not interact at all.  
However, even in the extreme case of non-interacting parties there is a possibility 
to correlate them.
Here we basically have two options: First, to perform a joint measurement on the two systems,
and second, to use another (ancillary) system. As the first option requires 
another system, the measurement apparatus, as well, we focus on 
the second approach where an additional system is used. 

If we look at the problem from the operational point of view we can solve 
the problem in the following way.
Let us suppose that we want to create an EPR pair shared by two parties
denoted as $A$ and $C$. We denote the additional party used to achieve 
the goal as $B$. First $B$ creates an EPR pair with party $A$ and a second EPR pair 
with party $C$. It means that the system $B$ is composed of two qubits.
Then by performing two-qubit (Bell) measurement on the two qubits at site $B$
we actually create an EPR pair shared by $A$ and $C$ irrespectively of the outcome
of the Bell measurement. 
The protocol outlined is the ``entanglement swapping'' protocol
proposed by M.~Zukowski et. al. \cite{Zukowski93}  and generalized by S.~Bose et. al.
\cite{Bose98}. The first experimental realization of the protocol
was done by J.~W.~Pan et.al. \cite{Pan98}.
\begin{figure}
\includegraphics[width=7cm]{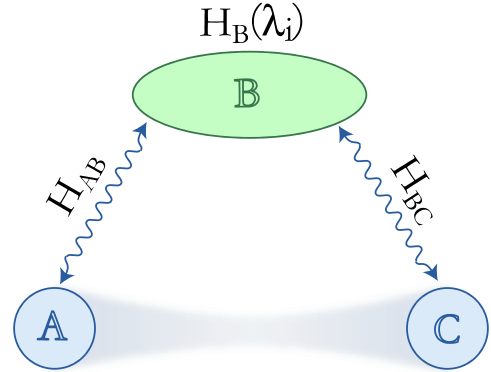}
\caption{Illustration of the physical situation.}
\label{obr2}
\end{figure}

In the entanglement swapping protocol instead of creating entanglement shared by $A$ and $C$
directly we create two maximally entangled pairs 
one shared by $A$ and $B$ and the second shared by $B$ and $C$. 
These entangled pairs can be produced  using interaction or joint measurement 
as we have discussed at the beginning. So the entanglement is created in the same 
way as before and only additional tools are used to transfer this entanglement into 
correlations between $A$ and $C$. 

In order not to use the same approach and explore different ways of creating the entanglement 
we modify the setup as follows.
The two systems $A$ and $C$ interact with the system $B$ and the interaction is described 
by the Hamiltonians $H_{AB}$ and $H_{BC}$. These Hamiltonians also include the
local terms $H_A$, and $H_C$. Let us remind that the systems $A$ and $C$ 
do not interact mutually and so the  Hamiltonian $H_{AC}$ is zero. 
In addition,  to control the entanglement produced between $A$ and $C$ 
we use the control of the system $B$ as is illustrated in Fig.~\ref{obr2}.

In such scenario we cannot assume that it is possible to create EPR pairs 
between $A$ and $B$ and $B$ and $C$. The result strongly depends
on the choice of the  Hamiltonians and if for example the mutual interaction is absent (the local
parts can be present though) no entanglement can be produced.
In this spirit it is an interesting question under which conditions it is possible to create 
quantum correlations between $A$ and $C$. 
It has recently been shown that for a large class 
of Hamiltonians if we monitor (measure) the system $B$ 
continuously \cite{Zhang03} or even non-continuously but repeatedly 
\cite{Verstraete04, Pachos04, Wu04} it is
indeed possible.

This result can be understood as follows. Let $t$ be the time 
of the free evolution of the system from the preparation to the measurement. 
If we prepare the three-partite system in a particular fully-factorized state, then 
after time $t$ the state of the system $ABC$ can be written as
\begin{eqnarray*}
| \psi (t) \rangle_{ABC} = \sum_j \: c_j(t)  \; 
|\phi_j (t) \rangle_{AC} \: | \omega_j (t) \rangle_B \; ,
\end{eqnarray*}
where $| \phi_j (t) \rangle_{AC}$ are vectors of unit length, $c_j$ 
are complex coefficients and 
 the basis $\{ | \omega_j (t) \rangle \}$ corresponds to the measurement that we perform 
at time $t$. This basis (or measurement) is chosen so that 
after performing the measurement and projecting the system onto 
one of the states $ |\phi_j (t) \rangle_{AC} \: | \omega_j (t)\rangle_B $ the corresponding state 
$| \phi (t) \rangle_{AC}$ of the subsystem $AC$ is entangled. \

Though the real protocol is more sophisticated, due to the probabilistic
nature of the measurement process, and we
need to monitor (measure) the system $B$ many times it 
 explains the main idea and the role of the measurement of the system $B$.
It is  the measurement that projects out the subsystem $AC$ onto an entangled state 
and the efficiency of detectors, incompleteness of the measurement itself or
complexity of the system $B$ can make the monitoring of the system $B$ difficult. 
As a consequence the entanglement is produced with a low degree or probability.
It is an open question whether it is possible to create 
entanglement between $A$ and $C$ without monitoring the system $B$. 
In such case the setup is the same as before (see Fig.~\ref{obr2})
but the only control that we have is a local ``coherent'' control of the system $B$. 
It means that we are allowed to  change the parameters of the Hamiltonian 
local at site $B$ but we are not allowed
to perform measurements. 
So what we can do is to ``drive'' the Hamiltonian and the system by changing the  
free parameters $\{ \lambda_i \}$ of the whole Hamiltonian $H$ of the three partite system $ABC$ 
\begin{eqnarray}
\label{ham}
H = H_{AB} + H_{BC} + H_B(\{ \lambda_i \}) \; ,
\end{eqnarray}
where $\{ \lambda_i \}$ are the parameters that correspond to 
the degrees of freedom that we control locally at site $B$.
Now the question is how much entanglement between $A$ and $C$ can be created 
by tuning the parameters $\lambda_i$.
The answer depends on the choice of the interaction Hamiltonians $H_{AB}$ 
and $H_{BC}$ and the local control at site $B$. 
In the following we discuss this dependence as well as the choice of different $B$'s.

The paper is organized as follows. In the next section we address
the case where all three systems $A$, $B$ and $C$ are two dimensional systems, 
that is qubits. We introduce the most general form of Hamiltonian consistent with
the assumptions and demonstrate the method on a particular example. 
In the third section we consider a more
complicated case of the Dicke model,  
where the role of the systems $A$ and $C$ are played by atoms interacting
with an electromagnetic field - the system $B$. Here different
approximations of the field are analyzed.  
In the last section we discuss various strategies as 
well as limits of the method and summarize our results.

\section{Qubits}
\label{sec2}

We start with the case where the systems $A$, $B$ and $C$ are represented by two-dimensional
Hilbert spaces and called qubits. In such case we can write the Hamiltonian $H_{AB}$ 
as a sum of direct products of Pauli matrices and the identity operator 
\begin{eqnarray}
\label{hab}
H_{AB} = \sum_{j,k=0}^3 h_{AB}^{jk} \; \sigma_A^j \otimes \sigma_B^k  \otimes \openone_C \; ,
\end{eqnarray}
where $\sigma_l^0$ is the identity operator $\sigma_l^0 = \openone$ and
$\sigma_l^j$, $j=1,2,3$ is the set of three Pauli matrices for each $l=A, B, C$.
It means that $\sigma_l^1 = \sigma_x$, $\sigma_l^2 = \sigma_y$ and $\sigma_l^3 = \sigma_z$.
In what follows we drop the subscript on operators labeling the system 
as the position of an operator in a product uniquely specifies to which system
the operator corresponds. Real constants, $h_{AB}^{jk}$, $j,k=0,..3$ define the 
interaction Hamiltonian $H_{AB}$. 
In the same way the real coefficients
$h_{BC}^{jk}$, $j,k=0,..,3$ uniquely define the Hamiltonian $H_{BC}$
\begin{eqnarray}
\label{hbc}
H_{BC} = \sum_{j,k=0}^3 h_{BC}^{jk} \; \openone \otimes \sigma^j  \otimes \sigma^k \; .
\end{eqnarray}

By local control on site $B$ we understand that we have a choice in tuning the
local Hamiltonian $H_B$ and more specifically parameters $h_B^j$ specifying the 
Hamiltonian 
\begin{eqnarray}
\label{hb}
H_B = \sum_{j=1}^3 h_B^j \; \openone \otimes \sigma^j \otimes \openone \; .
\end{eqnarray}
Note that in this case we did not include the case $j=0$ as such term only shifts 
energy levels but it does not essentially change the structure of the 
spectrum and eigenvectors.  
It means that the set of parameters $\{ \lambda_i \}$ we control 
are identified  with the three parameters $h_B^j$, $j=1,2,3$. 
To see how it works let us consider the following example. 

\subsection{Ising Interaction}

The Ising interaction between the sites $A$ and $B$ is described by the Hamiltonian \cite{ising}
\begin{eqnarray}
\label{izab}
H_{AB} = J \; \sigma^3 \otimes \sigma^3 \otimes \openone \; ,
\end{eqnarray}
where $J$ is the interaction constant and $\sigma^3 = \sigma^z$ is the
Pauli operator. Recalling the notation introduced above
we obtain that $h_{AB}^{33} = J$ and all other coefficients 
$h_{AB}^{jk}$, $j,k \neq 3$ are zero. 
The interaction between $B$ and $C$ is chosen to be the same
as it is the interaction between $A$ and $B$ and the Hamiltonian $H_{BC}$ reads
\begin{eqnarray}
\label{izbc}
H_{BC} = J \; \openone \otimes \sigma^3 \otimes \sigma^3 \; .
\end{eqnarray}
The physical situation that could be described by the interaction 
Hamiltonians is following.
Assume that the three systems $A$, $B$ and $C$ are of the same type. 
In such case also the interaction between $A$ and $B$ or 
 $B$ and $C$ is of the same origin. Subsequently if the three systems
are positioned in a line and equally spaced then the interaction between 
$A$ and $C$ is small, in practice negligible, while the interaction between $A$ and $B$ and the interaction between $B$ and $C$ 
are described by the same Hamiltonian as it is in our case.

The local Hamiltonians $H_A$ and $H_C$ for the Ising model are of the form
\begin{eqnarray}
H_A & = & \frac{\Delta}{2} \; \sigma^1 \otimes \openone \otimes \openone \; ,\\
H_C & = & \frac{\Delta}{2} \; \openone \otimes \openone \otimes \sigma^1 \; ,
\end{eqnarray}
where $\Delta$ is the energy separation between the upper and lower
energy level of the two-level system  and $\sigma^1=\sigma^x$ 
is the Pauli operator. For spin systems the parameter $\Delta$ is 
proportional to the magnitude of the 
external magnetic field that is responsible
for the splitting of the two energy levels when the spin
is placed in the magnetic field.
Similarly, the Hamiltonian $H_B$ 
is given by
\begin{eqnarray}
\label{izlb}
H_B = \frac{\Lambda}{2} \openone \otimes \sigma^1 \otimes \openone \; ,
\end{eqnarray}
where $\Lambda$ is the energy separation between the two energy levels of the 
system $B$. 
For convenience we rewrite the parameters $\Delta$ and $\Lambda$ as 
$\Delta \rightarrow \delta \Delta_0$ and  $\Lambda \rightarrow \lambda \Lambda_0$.
Here $\Delta_0$ and $\Lambda_0$ are chosen to be $\Delta_0 = \Lambda_0 = J$ and  
 $\lambda$ and $\delta$ are dimensionless. 
In this notation, the parameter that represents the control that we have 
over the system $B$ is the parameter $\lambda$. 
Change in the parameter $\lambda$ corresponds to the change of the strength of the external field 
at site $B$ which determines the energy separation between 
the two levels of the system $B$.

In what follows we discuss different choices of local Hamiltonians $H_A$ and $H_C$
and the Hamiltonian $H_B$ corresponding to the local control at site $B$. 
We start with the case where the local terms $H_A$ and $H_C$ are zero which means that
the parameter $\delta=0$.
The full Hamiltonian of the system $H$ is a sum of three terms (\ref{izab}), (\ref{izbc}) 
and (\ref{izlb}) and the corresponding eigenspectrum can be calculated 
analytically (see App.~\ref{apiz}). 
Due to  all of the eigenvectors are factorized it follows
immediately that there is no entanglement between $A$ and $C$
irrespective of the local Hamiltonian $H_B$. That is by locally controlling
 the system $B$ it is not possible to create entanglement shared by $A$ and $C$.
Note that the degeneracy of the ground state opens a chance to create
a state with the lowest energy such that it is entangled. 
However, as we can create an eigenstate with the same energy but no entanglement,
we will not consider such vector as an entangled ground state.	

A similar situation occurs when the local Hamiltonians $H_A$ and $H_C$ are large
(large means dominant compared with the interaction terms) so that the parameter
$\delta$ is much larger than $1$, that is $\delta \gg 1$. 
Large energy separation between two levels of systems $A$ and $C$ causes
that both of the systems $A$ and $C$ tend to be in their ground state. It follows that
the state of the subsystem $AC$ is close to a product of the two ground states, and hence
there is no entanglement between $A$ and $C$. Modifying $H_B$ cannot change this as the 
interaction terms are weak compared with the local Hamiltonians $H_A$ and $H_C$.

\begin{figure}[h]
\begin{center}
\includegraphics[width=8.5cm]{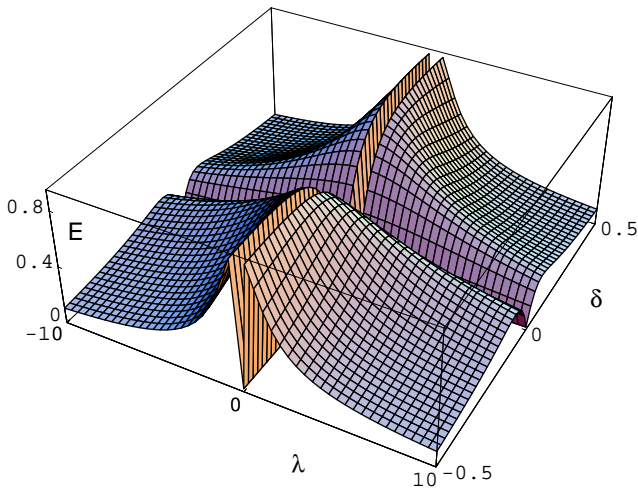}
\end{center}
\caption{Dependence of the entanglement $E$
shared by $A$ and $C$, expressed as a concurrence (for definition see App.~\ref{concu}),
on the local control at site $B$, parameter $\lambda$, and for different one-particle
Hamiltonians $H_A$ and $H_C$, parameter $\delta$.}
\label{graf2}
\end{figure}

On the other hand when the local terms are comparable to the interaction terms or even better when
they are small but non-zero the situation changes significantly. 
First let us consider  the local terms $H_A$ and $H_C$ to be small 
(but non-zero) in comparison with the interaction terms. 
It means that the parameter $\delta$ fulfills the relation $\delta \ll 1$.
In such case we can consider the local terms to be a perturbation to the 
full Hamiltonian and calculate the energy levels using the expansion series. 
As we have already pointed out 
for the Ising model without the local terms $H_A$ and $H_B$ the ground state
is degenerate and the two levels with the lowest energy are
\begin{eqnarray*}
|g_1 \rangle & = & | 0 \rangle \otimes | \alpha_1 \rangle \otimes | 0 \rangle \; ,\\
|g_2 \rangle & = & | 1 \rangle \otimes | \alpha_7 \rangle \otimes | 1 \rangle \; ,
\end{eqnarray*}
for more details see App.~\ref{apiz}. If we include the local terms $H_A$ and $H_C$ in the full 
Hamiltonian we remove the degeneracy and the ground state becomes non-degenerate. 
Calculating the ground state to the first order in the parameter $\delta$ with the help 
of the expansion to the second order  (the first order neither removes the degeneracy 
nor modifies the spectrum but modifies the form of the two ground states) 
we obtained  the state of the form
\begin{eqnarray*}
1/K \; \left \{ \;   |g_1 \rangle + | g_2 \rangle + \delta \; [ \; | 0 \rangle \otimes 
(c_1 | \alpha_3 \rangle + c_2 | \alpha_4 \rangle ) \otimes | 1 \rangle  \right . \\
\left . +  | 1 \rangle \otimes 
(c_3 | \alpha_3 \rangle + c_4 | \alpha_4 \rangle ) \otimes | 0 \rangle  \;  ] \; \right \}  \; ,
\end{eqnarray*}
where the complex constants $c_j$, $j\!=\!1,..,4$ depend on $\lambda$, $K$ is a
normalization constant and the normalized states $| \alpha_j \rangle$ are defined in App.~\ref{apiz}. 
For $\delta$  small this vector represents an entangled state
and for $\delta \! \rightarrow \! 0$ the ground state approaches maximally entangled state.
Though not explicitly shown in the last equation 
change in the parameter $\lambda$ causes change in the state (as not only the constants $c_j$ 
but also the states $| \alpha_j \rangle$  depend on $\lambda$) and in turn changes the
entanglement shared by $A$ and $C$. 
On the other hand it should be pointed out that the maximal entanglement can be reached only in the limit 
$\delta \rightarrow 0$ and in the same limit the gap between the ground state and the first
excited state vanishes. It means that if we want to increase the maximal amount of entanglement
that can be produced between $A$ and $C$ we have to reduce the gap between the two lowest 
energy levels. 
As a result of that the cooling of  the system (we want our system to 
stay in the ground state during the whole evolution) is more problematic. 
The complete picture for arbitrary values of the parameters $\delta$ and $\lambda$ is
shown in Fig.~(\ref{graf2}).

Using simple interaction of Ising type 
  we have shown that it is possible to control (generate) entanglement between
essentially distant parties $A$ and $C$. What is important to realize 
is the fact that the two systems $A$ and $C$  are not allowed to interact and the entanglement 
 is created only through the interaction with the system $B$. In addition, by modifying the 
site $B$, that is parameters of the local Hamiltonian $H_B$, it is possible to control the amount of 
entanglement shared by $A$ and $C$. 
In our example the maximal entanglement is actually never reached though we can get arbitrarily
close to the maximal possible value. We address this issue in a more detail in the last section.

\section{Atoms interacting with a single mode electromagnetic field}

In our scenario  the entanglement between the two parties $A$ and $C$ \
is generated through the interaction 
with an auxiliary quantum system at the site $B$ 
and depends on the choice of the local Hamiltonian $H_{B}$.
That is it depends on the physical nature of the system $B$. 
In the previous section the system $B$ was composed of a single qubit. 
The situation analyzed in Ref. \cite{Bose03} can be considered 
as a particular case of our scenario  
where the system $B$ is a collection of spins and the whole system $ABC$
forms a spin chain. 
It is natural that for different physical systems $B$ we obtain 
different results.
What is not so obvious is the fact that different results are obtained also 
for different models of a given physical system. 
Here by different models we have in mind different approximations of 
the physical situation. 

In order to see the problem more clearly  
we analyze the physical setup composed of two non-interacting atoms 
placed in a cavity interacting with one mode 
electromagnetic field. 
As the two atoms do not interact directly the entanglement 
can be created only via interaction with the electromagnetic
field. Here different approximations lead to different models 
for the field and one can consider several Hamiltonians.

If we assume a dipole and rotating wave approximation (RWA) 
and restrict to the case of small interaction between the field and 
the atomic system, then the system can 
be described by Dicke Hamiltonian \cite{Dicke1954}.
\[
H_1(\kappa) =\frac{\omega _{A}}{2}\sum_{j}\sigma
_{j}^{3}+\omega _{F}a^{\dagger }a +\kappa \sum_{j}(\sigma _{j}^{+}a+\sigma _{j}^{-}a^{\dagger }) \; ,
\]
where the  $\sigma _{j}^{+}=\frac{1}{2}(\sigma _{j}^{1}+i\sigma _{j}^{2})$, 
$\sigma _{j}^{-}=\frac{1}{2}(\sigma _{j}^{1}-i\sigma _{j}^{2})$, the
three operators $\sigma ^{1}$, $\sigma ^{2}$,$\sigma ^{3}$
 are Pauli operators, $a^{\dagger}$ and $a$ are field creation and 
annihilation operators, $\omega_F$ is the radiation field frequency 
and $\omega_A$ is the atomic transition frequency. 
The parameter $\kappa$ is proportional to the coupling strength 
between field and atoms.

The Dicke Hamiltonian is a good starting point in the analysis of 
the radiation-matter interaction systems, since it can be 
analytically solved \cite{Tavis1967} and describes 
various (especially collective and cooperative) properties of 
these systems \cite{Wolf}.

Nevertheless, if we want to study the system with 
stronger radiation-atom interaction 
we must drop the rotating wave-approximation (i.e. it is 
important to include 
counter-rotating terms to the Hamiltonian) 
and in addition an extra quadratic field term, usually neglected, 
have to be taken into account. 
When the counter-resonant terms are added the Hamiltonian is of the
form:
\begin{eqnarray}
H_2(\kappa) &=&\frac{\omega _{A}}{2}\sum_{j}\sigma
_{j}^{3}+\omega _{F}a^{\dagger }a+  \nonumber \\
&+&\kappa \sum_{j}(\sigma _{j}^{+}a+\sigma _{j}^{-}a^{\dagger }+\sigma
_{j}^{+}a^{\dagger } + \sigma _{j}^{-}a)  \; ,\nonumber
\end{eqnarray}
and when in addition the quadratic field term is included the Hamiltonian 
reads(for more details see Ref.~??):
\begin{eqnarray}
H_3(\kappa ,\lambda ) &=&\frac{\omega _{A}}{2}\sum_{j}\sigma
_{j}^{3}+\omega _{F}a^{\dagger }a+  \nonumber \\ \label{eq:dicke3}
&+&\kappa \sum_{j}(\sigma _{j}^{+}a+\sigma _{j}^{-}a^{\dagger }+\sigma
_{j}^{+}a^{\dagger } + \sigma _{j}^{-}a) \\
&+&\lambda (a+a^{\dagger })^{2} .  \nonumber
\end{eqnarray}
In both of the Hamiltonians $H_2$ and $H_3$ the parameter 
$\kappa $ denotes the coupling between field 
and matter and the parameter $\lambda$ that appears in 
$H_3$ is not a free parameter but it is proportional to $\kappa^2$ 
(for more details see Ref.~??).

The quadratic term in the Hamiltonian (\ref{eq:dicke3}) is not only 
necessary for the cases of stronger 
interactions from the physical point of view, but it is also useful 
for the mathematical analysis, how an
 extra non-interaction term included into the Hamiltonian influences 
the properties of the system. 
\begin{figure}
\begin{center}
\includegraphics[width=8cm]{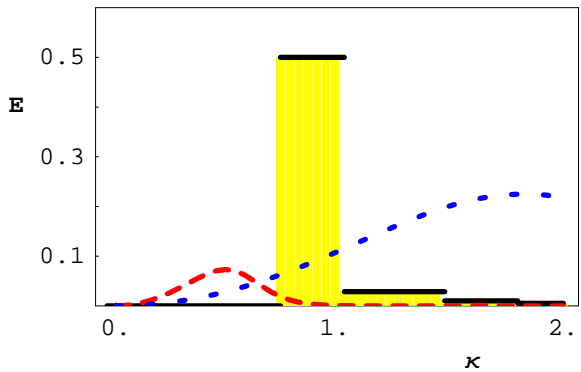}
\end{center}
\caption{Concurrence $E$ of the two atoms in the cavity  as a function of the coupling $\kappa $ for three
 different types of Hamiltonians $H_1$ (black solid line), $H_{2}$ (red dashed line) and $H_{3}$
 (blue dashed line). The behavior of the bi-partite entanglement in the two atom system is 
 clearly different for the three Hamiltonians.
 The amount of quantum correlation in the system depends on and thus can be controlled by 
the strength of the interaction.
 It is apparent from the figure that for the case of Hamiltonian $H_3$ the concurrence persists also in
 the cases of  stronger interactions and in average it has larger values compared with cases of
$H_1$ and $H_2$.}
\label{fig:entanglement1}
\end{figure}

In what follows, we will study how the change of the parameters $\kappa$ and $\lambda$ influences the
 entanglement between individual two particles of the atomic system. More specifically, we will study
 bi-partite atomic entanglement in the  ground state of the systems described by the Hamiltonians $H_1$,
 $H_2$ and $H_3$. Finally, we will compare the results.  Since atoms are described as two-level quantum 
systems, we will use the concurrence of a reduced bipartite atomic system as an entanglement measure.

Firstly, we focus on how the quadratic field term in Hamiltonian $H_3$ influences the bi-partite
 entanglement, comparing to the Hamiltonians without this term ($H_1$ and $H_2$). 
From all tree Hamiltonians, only Hamiltonian $H_1$ can be diagonalized analytically
 \cite{Tavis1967, Buzek2005}, therefore we have done an numeric analysis and we will discuss the 
 results on figures. It is apparent from the Fig.~\ref{fig:entanglement1} that the bi-partite atomic
 entanglement is significantly different in the three cases studied.  The main property of the concurrence
 for the Hamiltonians $H_1$ and $H_2$ (i.e. without non-interacting term) is that their values drop
 to $E = 0$ very quickly. On the other hand the presence of the quadratic non-interacting field term
 in $H_3$ ensures that the quantum correlations persist also in the cases of strong interaction
 between field and atomic system. In addition, two atoms are (in average) more
 entangled compared with the previous cases as it can be seen from the Fig.~\ref{fig:entanglement1}. 

\begin{figure}
\begin{center}
\includegraphics[width=8cm]{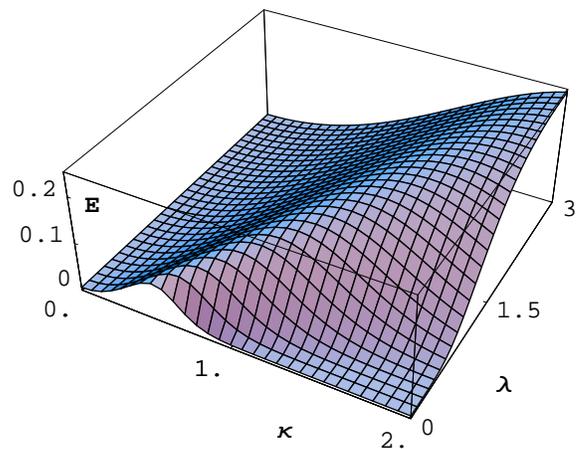}
\end{center}
\caption{Bipartite entanglement $E$ (concurrence) between the two atoms as a function of the
 coupling $\kappa$ and the parameter $\tilde{\lambda}$. It is apparent from the figure that we can control
 (increase) the bipartite entanglement by increasing the system interaction $\kappa$ or by
 increasing the quadratic term independently  (the parameter $\tilde{\lambda}$).}
\label{fig:entanglement2}
\end{figure}

Further, let us separately study how the bi-partite entanglement in the atomic system is controlled
 by the strength of the  interaction (parameter $\kappa$) and the size of the quadratic term
 (parameter $\lambda$) in the case of the complete Hamiltonian $H_3$. We note that this is rather
 mathematical approach since $\lambda$ is not independent from the coupling ($\lambda \sim \kappa^2$)
 but it can illuminate how the non-interacting field term can have an influence on the entanglement
 between individual atoms - even for fixed coupling $\kappa$.
Again, all calculations were made numerically and our results will be discussed with the help of the
 Fig.~\ref{fig:entanglement2}. 

As it is illustrated in the figure, by the change in the parameters in the quadratic field term 
($\lambda \rightarrow  \tilde{\lambda} \lambda$ ) 
we can control the strength of the bi-partite quantum 
correlations in the atomic system. These get stronger as the interaction between field and atomic
 system increases (increasing  $\kappa$) or when we independently increase the quadratic term
 (increasing $\tilde{\lambda}$).

\section{General case}

In this section we address the limits of the studied scheme and for that 
we introduce the following theorem.
\mbox{} \\

\noindent
{\em Theorem:} Consider Hamiltonian (\ref{ham}) that is symmetric under exchange of
the labels (systems) $A$ and $C$. Further let $| \psi \rangle_{ABC}$ be an eigenstate
of the Hamiltonian $H$ such that it is of the form 
$ | \psi \rangle_{ABC} = | \omega \rangle_{AC} \otimes | \beta \rangle_B$.

If the energy level corresponding to the state $| \psi \rangle_{ABC}$ is non-degenerate
then the state $| \psi \rangle_{ABC}$ is fully factorized so that 
$| \omega \rangle_{AC} = | \alpha \rangle_A \otimes | \gamma \rangle_C$. 
\mbox{} \\

\noindent
{\em Proof:} First we rewrite the state $| \psi \rangle_{ABC}$ so that we decompose the 
state $| \omega \rangle_{AC}$ using the Schmidt basis
\begin{eqnarray}
\label{stsch}
|\psi \rangle_{ABC} = \sum_j \sqrt{\lambda}_j | j \rangle_A \otimes | \beta \rangle_B \otimes
| j \rangle_C \; ,
\end{eqnarray}
where $\lambda_j$ are eigenvalues of the density operator corresponding to the
state $\rho_A = {\rm Tr}_C (| \omega \rangle \langle \omega |)$ and $\{ | j \rangle_A \}$ and $\{ | j \rangle_C \}$ 
is the Schmidt basis of the system $A$ and $C$ respectively.
Applying Hamiltonian (\ref{ham}) to the state (\ref{stsch}) we obtain the expression
\begin{eqnarray*}
H | \psi \rangle_{ABC} = \sum_j \sqrt{\lambda}_j [ | w_j \rangle_{AB} \otimes | j \rangle_C
 + | j \rangle_A \otimes | w_j \rangle_{BC} ] \; ,
\end{eqnarray*}
where $| w_j \rangle_{AB} = (H_{AB} + H_B/2) | j \rangle_A \otimes | \beta \rangle_B$, 
$| w_j \rangle_{BC} = (H_{BC}+H_B/2) | \beta \rangle_B \otimes | j \rangle_C$.
The action of the Hamiltonian $H$ was divided into two parts: the interaction between
$A$ and $B$ and the interaction between $B$ and $C$. In order to preserve symmetry under the
exchange of $A$ and $C$ the term corresponding to the local control at site $B$ was
divided into two equal parts. One of them was added to $H_{AB}$ and the other to 
$H_{BC}$. Notice that the states $w_j$ need to be neither normalized nor mutually orthogonal.
For the state $| \psi \rangle_{ABC}$ to be an eigenstate of the Hamiltonian the reduced operator 
of the system $B$ has to be proportional  to the projection $| \beta \rangle_B \langle \beta |$ 
(in the propositions of the theorem we assume a particular form of the state
$ | \psi \rangle_{ABC}$). It follows that the action of the Hamiltonian 
is restricted~\footnote{In order to obtain the result it is crucial 
that the Hamiltonian $H$ is invariant under the exchange of the labels (systems) $A$ and $C$.} and
\begin{eqnarray*}
(H_{AB} + H_B/2)  | j \rangle_A \otimes | \beta \rangle _B & = &
| v_j \rangle_A \otimes  | \beta \rangle_B \; , \\
(H_{BC} + H_B/2)  | \beta \rangle _B \otimes | j \rangle_C & = &
| \beta \rangle_B \otimes | v_j \rangle_C  \; , 
\end{eqnarray*}
where the vectors $| v_j \rangle$ are unnormalized in general.
In addition, the state $H | \psi \rangle_{ABC}$  has to be orthogonal to the state of the form
$ | j \rangle_A \otimes | \beta \rangle_B \otimes | k \rangle_C$ where $k \neq j$.
It follows that $\langle k | v_j \rangle = 0$ for $k\neq j$ and the action of $H_{AB}+H_B/2$ 
(correspondingly $H_{BC}+H_B/2$) is of the form
\begin{eqnarray*}
(H_{AB} + H_B/2)  | j \rangle_A \otimes | \beta \rangle_B \otimes | j \rangle_C = 
c_j | j \rangle_A \otimes | \beta \rangle_B \otimes | j \rangle_C \; .
\end{eqnarray*}

Finally, if the sum over $j$ in (\ref{stsch}) includes only a single term then the state 
$| \psi \rangle_{ABC}$ 
is fully factorized. On the other hand if the sum includes two and more terms 
then the constants $c_j$ for that  terms have to be equal (independent of $j$). 
However in such case the energy level corresponding to that state is degenerate. 
It follows from the fact that using the expressions for $H_{AB}$ and $H_{BC}$ derived above 
it is possible to show that the states of the form 
$\sum_j c_j |j \rangle_A \otimes | \beta \rangle_B \otimes | j \rangle_C$, where $c_j$ 
are arbitrary complex numbers up to normalization, have the same energy.
\mbox{} \\

\noindent
The Theorem has important implications concerning the creation of entanglement between $A$ and $C$.
It follows from the theorem that using the method it is not possible to create a maximally
entangled state shared by $A$ and $C$. More specifically, 
if the ground state of the system $ABC$ is such that the reduced state of $AC$ 
is a maximally entangled state then we know from the theorem that the ground state is 
degenerate and it is possible to create a non-entangled state with the same energy so 
we should not consider such ground state as entangled. On the other hand we can be
arbitrarily close to a maximally entangled state and this was demonstrated in the Sec.~\ref{sec2}.

Further, applying the theorem more generally we can state that in this scenario 
it is not possible to create any pure 
entangled state shared by $A$ and $C$. It means that by modifying the local parameters 
at site $B$ and not considering measurements the ground state of the system 
$ABC$ is such that the reduced state $\rho_{AC}$ of the systems $A$ and $C$ 
can be entangled only if it is mixed. 

To summarize, we have analyzed a particular scenario of the generation of entangled
where the entanglement is produced via interaction with additional
system and no measurements are considered. We have shown that under the symmetry condition
the maximal entanglement can be reached only asymptotically and no pure entangled state
can be produced. Moreover, we have shown that different assumptions about the 
additional physical system $B$ result into situations where 
different amount of entanglement is produced. 


\begin{center}
{\bf Acknowledgement} 
\end{center}

\noindent 
This work was supported in part by the
European Union projects INTAS-04-77-7289, CONQUEST and QAP,  by
the Slovak Academy of Sciences via the project CE-PI/2/2005, by the
project APVT-99-012304.

\appendix
\section{Ising model}
\label{apiz}

In this appendix we present the eigenvectors and corresponding 
eigenvalues of the Ising Hamiltonian (\ref{izab}) without the 
local terms at sites $A$ and $B$ but 
with the most general local term at site $B$. The Hamiltonian $H$ with the
Ising type interaction between sites $A$ and $B$ and sites $B$ and $C$ together with 
the most general local term corresponding to the site $B$ is of the form 
\begin{eqnarray*}
H & = & \sigma^3 \otimes \sigma^3 \otimes \openone + \openone \otimes \sigma^3 \otimes \sigma^3 \\
& & + \openone \otimes ( \sum_j h_B^j \sigma^j ) \otimes \openone \; .
\end{eqnarray*}
The eigenvectors of the Hamiltonian $H$ with the corresponding eigenvalues are listed below
\begin{eqnarray*}
\begin{array}{lll}
e_{1,2} = | 0 \rangle \otimes | \alpha_{1,2} \rangle \otimes | 0 \rangle, & E_{1,2} = \pm \sqrt{\sum_j (h_B^j+v^j)^2} \; ,\\
e_{3,4} = | 1 \rangle \otimes | \alpha_{3,4} \rangle \otimes | 0 \rangle, & E_{3,4} = \pm \sqrt{\sum_j (h_B^j)^2} \; ,\\
e_{5,6} = | 0 \rangle \otimes | \alpha_{5,6} \rangle \otimes | 1 \rangle, & E_{5,6} = \pm \sqrt{\sum_j (h_B^j)^2} \; , \\
e_{7,8} = | 1 \rangle \otimes | \alpha_{7,8} \rangle \otimes | 1 \rangle, & E_{7,8} = \pm \sqrt{\sum_j (h_B^j-v^j)^2} \; , \\
\end{array}
\end{eqnarray*}
where $v^j = (0,0,2)$, and the vectors $| \alpha_j \rangle$, $j=1,2$  are two eigenvectors 
of the operator $h_B^j \sigma^j + 2 \sigma^3$, the vectors $| \alpha_j \rangle = | \alpha_{j+2} \rangle$, 
$j=3,4$ are eigenvectors of the operator $\sum_j h_B^j \sigma^j$ and the vectors $| \alpha_j \rangle$, 
$j=7,8$ are eigenvectors of the operator $\sum_j h_B^j \sigma^j - 2 \sigma^3$.
\mbox{} \\

\section{Concurrence}
\label{concu}

In this appendix we recall the definition of the concurrence \cite{Wootters97} which is a measure of
bipartite entanglement shared by two qubits (quantum systems associated to two-dimensional
Hilbert spaces). Let $\rho_{AB}$ be a bipartite state (density matrix) of a two-qubit system.
Further, denote as $\lambda_i$, $i=1,2,3,4$ the eigenvalues of the matrix 
$\rho_{AB} \; \sigma^2 \otimes \sigma^2 \rho_{AB}^* \; \sigma^2 \otimes \sigma^2$ 
listed in a non-decreasing order. Here $\rho_{AB}^*$ means complex conjugation of the 
matrix $\rho_{AB}$ and $\sigma^2$ is the Pauli operator corresponding to the 
measurement of the spin along the $y$ axis. Then the concurrence $E$ is defined as
\begin{eqnarray}
E (\rho_{AB}) \equiv {\rm Max} 
\{0, \sqrt{\lambda_1} \! - \! \sqrt{\lambda_2} \! - \! \sqrt{\lambda_3} \! - \! \sqrt{\lambda_4} \} .
\end{eqnarray}


\end{document}